%% file: QOC_QSL.tex
\documentclass[prx,twocolumn,nofootinbib,floatfix,10pt]{revtex4-1}
\usepackage[pdftex]{graphicx} 
\usepackage{hyperref}

\usepackage[utf8]{inputenc}
\usepackage{amsfonts, amsmath, amssymb, bm, nccmath}
\usepackage{siunitx}
\usepackage{import}
\DeclareSIUnit\Er{\ensuremath{E_{\text{r}}}}
\newcommand\Er{\si{\Er}}
\usepackage[british]{babel}
\usepackage{color}

\usepackage[normalem]{ulem}
\usepackage{lpic}
\usepackage{textcomp}

\begin{document}

\newcommand{\ket}[1]{|{#1}\rangle}
\newcommand{\bra}[1]{\langle{#1}|}
\newcommand{\braket}[2]{\langle{#1}|{#2}\rangle}

\title{Optimal control of complex atomic quantum systems}

\author{ S. van Frank$^{1}$, M. Bonneau$^{1}$, J. Schmiedmayer$^1$,\\
S. Hild$^{2}$, C. Gross$^2$, M. Cheneau$^{2,3}$,  I. Bloch$^2$,\\ 
T. Pichler$^{4}$, A. Negretti$^5$, T. Calarco$^4$, S. Montangero$^4$}

\affiliation{$^1$Vienna Center for Quantum Science and Technology, Atominstitut, TU Wien, Stadionallee 2, A-1020 Vienna, Austria}

\affiliation{$^{2}$Max-Planck-Institut f\"{u}r Quantenoptik, Hans-Kopfermann-Str.~1, 85748 Garching, Germany}

\affiliation{$^{3}$Laboratoire Charles Fabry, Institut d’Optique - CNRS - Université Paris Saclay, 91127 Palaiseau}

\affiliation{$^{4}$Institute for complex quantum systems \& Center for Integrated Quantum Science and Technology (IQST), 
Universit\"at Ulm, Albert-Einstein-Allee 11, D-89075 Ulm, Germany}

\affiliation{$^{5}$Zentrum f\"ur Optische Quantentechnologien and The Hamburg Centre for Ultrafast Imaging, Universit\"at Hamburg, Luruper Chaussee 149, D-22761 Hamburg, Germany}


\begin{abstract}
{\bf 

Quantum technologies will ultimately require manipulating many-body quantum systems with high precision. Cold atom experiments represent a stepping stone in that direction: a high degree of control has been achieved on systems of increasing complexity, however, this control is still sub-optimal. Optimal control theory is the ideal candidate to bridge the gap between early stage and optimal experimental protocols, particularly since it was extended to encompass many-body quantum dynamics.  
Here, we experimentally demonstrate optimal control applied to two dynamical processes subject to interactions: the coherent manipulation of motional states of an atomic Bose-Einstein condensate and the crossing of a quantum phase transition in small systems of cold atoms in optical lattices. {We show theoretically that these transformations can be made} fast and robust with respect to perturbations, including temperature and atom number fluctuations, resulting in a good agreement between theoretical predictions and experimental results. 
}
\end{abstract}

\maketitle


The last two decades have seen exceptional progress in the ability to engineer, manipulate and probe complex quantum systems. The concepts of quantum computation, quantum simulation or precision measurement beyond the classical limit have been validated in the laboratory and quantum sensing and metrological devices have been developed for specific applications~\cite{rosi_precision_2014, poli_precision_2011, neumann_2013, bloom_2014,bloch_2012,georgescu_2014}. Another important challenge to meet in order to fully exploit the potential of complex quantum systems is to design more robust and efficient experimental protocols.

Most of the protocols developed so far in research laboratories rely on {analytic or simple empirical} solutions. {In the paradigmatic example of a superfluid-to-Mott-insulator transition in a lattice, adiabatic manipulations are generally applied. Although maintaining adiabaticity is impossible in the thermodynamic limit, almost adiabatic transformations can become feasible for finite size systems. However, they are -- by definition -- slow compared to the typical timescales of the system. Thus, they are highly sensitive to decoherence and experimental imperfections. Speeding up the transformation can lead to a significant gain in that regard. 
In another common case, the driving of a transition {between two energy levels} of a system, an adiabatic solution does not necessarily exist. The transition can, under certain constraints, be driven by a Rabi pulse at the frequency of the level splitting. However, in the presence of other accessible levels or loss processes, this option has a strongly limited efficiency. It would therefore be desirable to have at our disposal a method to design fast and arbitrary complex manipulations. In addition, to be experimentally sound, such a method would have to be robust with respect to system perturbations.}
\begin{figure*}[t]
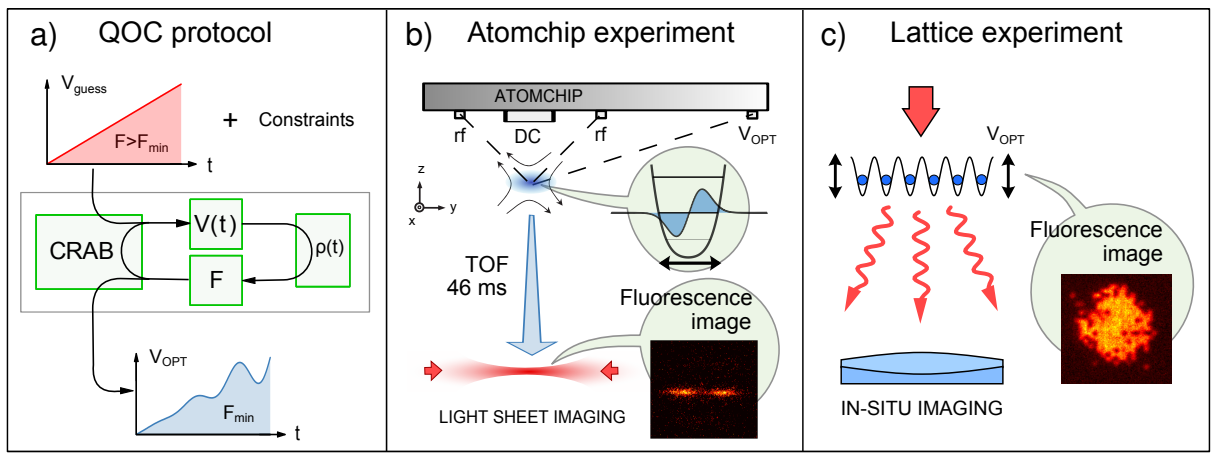

	\centering
\begin{lpic}[l(0mm),r(0mm),t(0mm),b(0mm)]{Fig1(160mm)}
\lbl[t]{6,77;{\fontfamily{phv} \selectfont \large a)}}
\lbl[t]{73,77;{\fontfamily{phv} \selectfont \large b)}}
\lbl[t]{147,77;{\fontfamily{phv} \selectfont \large c)}}
\end{lpic}
	\caption{a) {The CRAB optimal control applies a first control field $V_\text{guess}(t)$ to a numerically simulated experiment. Taking into account experimental constraints, it optimizes the control field relying on the figure of merit $F$ after time evolution. The final control field obtained after optimization, $V_\text{OPT}(t)$, optimally steers the system in the minimal possible time $T_\text{OPT}$ compatible with the theoretical and experimental limitations.}\\
	b) Vienna atomchip experimental setup: illustration of the experimental setup with the atomchip (top) used to trap and manipulate the atomic cloud (middle) and the light sheet as part of the imaging system (bottom). The trapping potential, centered on a DC wire, is made slightly anharmonic by alternating currents in the two radiofrequency (rf) wires. It is then displaced (black arrow) along the optimal control trajectory $V_\text{OPT}(t)$, using an additional parallel wire located far away from the DC wire and carrying a current proportional to $V_\text{OPT}(t)$. By this mechanical displacement of the wavefunction, a transition is realized from the ground to the first excited state of the trap. The atomic cloud is imaged after a $\SI{46}{\milli\second}$ time-of-flight. \\
	c) 
	Garching lattice experiment setup: an optical lattice is applied along an array of tubes and drives the superfluid to Mott-insulator transition with one atom per site (top) {following the optimized control field $V_\text{OPT}(t)$ (black arrows)}. The distribution of atoms in the Mott regime is probed by fluorescence imaging through a high-resolution microscope objective with single-site resolution and single-atom sensitivity (bottom right).}
	\label{fig:ls}
\end{figure*}
This challenge can be met by means of optimal control theory, that is, the automated search of an optimal control field to steer the system towards the desired goal~\cite{kirk_optimal_2004,brif_2010}. Quantum Optimal Control (QOC) has been applied very successfully in the case of (effective) few-body quantum systems: it has been shown that QOC can steer the dynamics in the minimum allowed time and that the optimal protocols are robust with respect to noise~\cite{brif_2010}. In particular, 
it has been experimentally demonstrated, 
for quantum dynamics taking place in an effective two-level system, that QOC allows to saturate the Quantum Speed Limit (QSL) -- the  minimal  time necessary to transform one state into another for a given energy of the driving~\cite{bhattacharyya_quantum_1983, margolus_maximum_1998, levitin_fundamental_2009, deffner_quantum_2013,del_campo_quantum_2013,bason_2012}. However, only recently QOC has been extended to embrace many-body quantum dynamics in non-integrable quantum systems~\cite{doria_2011,rosi_2013,caneva_2012,lloyd_2014,burgarth_2010}.


In this paper, we apply QOC to two ultracold atom systems undergoing complex transitions in the non-perturbative regime and we show that it is possible to speed up their dynamics at timescales comparable with the QSL theoretical estimate (Fig.~\ref{fig:ls}a). {The two selected experiments are prime examples of quantum systems in which the dynamics is affected by interactions between particles and where much is to be gained from speeding up the attempted transformation.}
{In the first experiment}, we demonstrate a fast control scheme for the motional state of a quasi-condensate on an atom chip (Fig.~\ref{fig:ls}b).  
The complexity of this system arises from the multiplicity of accessible motional states and the non-linearities induced by atom-atom interactions. {We show that, using a mean-field Gross-Pitaevskii representation of the system, QOC is successful at optimizing state transformations or state preparation, making it a versatile tool for potential quantum information processing applications}. Following an optimized non-trivial trajectory, we achieve theoretical and experimental infidelities below 1\% on a timescale {on the order of the QSL.} 
{In the second experiment}, we experimentally speed up the crossing of the one-dimensional superfluid Mott-insulator (SF-MI)  quantum phase transition of cold atoms in an optical lattice (Fig.~\ref{fig:ls}c). This experiment {is the first example of QOC applied to quantum phase transitions {and might have implications to improve the efficiency of future adiabatic quantum computation protocols}.} It also shows the power of QOC to efficiently treat many-body dynamics: the optimal quantum phase transition crossing is about one order of magnitude faster than the adiabatic protocol while maintaining the same final state fidelity.

As we will show, the optimal transformations we engineer are in general robust with respect to moderate fluctuations of the systems' parameters, finite temperature and atom number fluctuations, a fundamental requirement for a successful application of optimal protocols to experimental systems.


\section*{Optimal control and quantum speed limit}

A typical optimal control problem is defined as follows: given a dynamical law 
which defines the time evolution of the system state $\rho$ and depends explicitly on an external control field $V(t)$, one looks for the optimal control field $V_\text{OPT}(t)$ such that a quantity of interest --- the {figure of merit} $F(V(t))$ --- is extremized (Fig.~\ref{fig:ls}a). In the following, among the different algorithms that have been successfully developed to perform QOC processes~\cite{brif_2010,dalessandro_2007}, we will exploit an approach recently introduced by some of us~\cite{doria_2011,caneva_chopped_2011}: the Chopped RAndom Basis (CRAB) optimization. This approach has been specifically designed to solve optimal control problems where the access to the knowledge of the system properties is limited and/or the computation of the figure of merit is highly demanding (see appendix A for details): for example when using tensor network methods~\cite{white_1992, schollwock_2011}, multi-configuration time-dependent Hartree Fock  methods~\cite{Alon2008,Cao2013,brouzos_2014}, or in a closed-loop setting whenever the optimization is performed directly as part of the experimental cycle~\cite{rosi_2013}.

Despite the successes of QOC, there are fundamental limitations that clearly cannot be overcome. One of the most fundamental ones is related to the energy-time uncertainty -- the QSL. It accounts for the fact that the system's finite energy defines a minimal time-scale needed for the system to react~\cite{bhattacharyya_quantum_1983}. The simplest instance of such fundamental limit is, in a two-level system, given by the Rabi frequency which provides a lower bound for the time needed to perform a transition between the levels. More generally, it can be proven that the time needed to perform a given transformation between two states is bounded by $T \ge d(\rho_i, \rho_f)/ \overline \Lambda$, where $d(\rho_i, \rho_f)$ is the distance between the initial and the final states, and $\overline{\Lambda}$ is the time-averaged p-norm of the Liouvillian~\cite{bhattacharyya_quantum_1983, margolus_maximum_1998, levitin_fundamental_2009, deffner_quantum_2013,del_campo_quantum_2013}. Whenever the previous bound is saturated, the system is said to be driven at the QSL~\cite{Caneva2009}. An independent heuristic estimate of the QSL can be provided by solving the (constrained) optimal control problem at different total times $T$ for fixed control field strength: the minimal time needed to reduce the figure of merit below some critical threshold can be defined as the QSL of the process. It has been shown in a few cases that these two definitions coincide~\cite{Caneva2009,caneva_speeding_2011,brouzos_2014}.

The functional dependence with the total time $T$ of the final figure of merit depends on the specific system considered: if the time-optimal transformation follows the geodesic in the Hilbert space between the initial and the final (orthogonal) state at maximum 
constant speed, the final figure of merit is given by {$F(T)/F(T=0)=\cos^2({\pi T}/(2 T_\text{QSL}))$}~\cite{Caneva2009,caneva_speeding_2011,brouzos_2014} {for $T \leq T_\text{QSL}$.} This simple formula allows then to estimate the QSL by means of a one-parameter fit according to the previous definition. It can be either directly applied or adapted, as we will show in the two optimal processes studied. Finally, when applying this approach to describe any experimental setup, one should take into account the deviations from the idealized theoretical model and the concrete measurement capabilities, which introduce a limited distinguishability between states (in terms of any measurable quantity): hereafter we choose as optimal total 
duration of our experiments $T_\text{OPT}<T_\text{QSL}$, the minimal time where the figure of merit reaches the minimum compatible with the experimental resolution.

In summary, in the following we perform a numerical CRAB optimization for different final times $T$ of two complex dynamical processes (illustrated in Fig. \ref{fig:ls}b-c), from which we obtain a theoretical estimate of $T_\text{OPT}$ and the corresponding optimal control fields, which we use successfully to experimentally manipulate the system at a timescale compatible with the QSL.
\begin{figure*}[t]
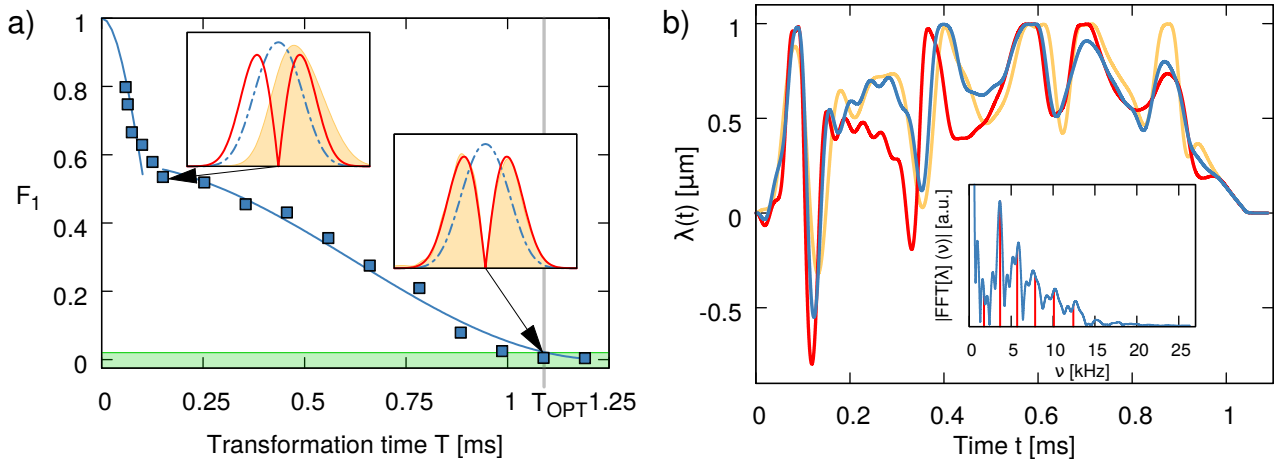

\centering
\begin{lpic}[l(0mm),r(0mm),t(0mm),b(0mm)]{Fig2a(86mm)}
\lbl[t]{2,60;{\fontfamily{phv} \selectfont \large a)}}
\end{lpic}
\begin{lpic}[l(0mm),r(0mm),t(0mm),b(0mm)]{Fig2b(86mm)}
\lbl[t]{2,60;{\fontfamily{phv} \selectfont \large b)}}
\end{lpic}
\caption{a) Theoretical prediction of the infidelity $F_1$ as a function of the transformation time $T$ achieved by optimal driving of the ground-to-first-excited state transfer for $N=700$ (blue squares).  The blue solid lines are fits of the numerical results according to $\alpha \cos^2(\beta T)$ with $\alpha \simeq 1, 0.56$ and $\beta \simeq 7.35, 1.18$ (left and right curve respectively). The green region represents the minimum measurable infidelity in our present experiment. 
Insets: modulus squared of the evolved wave function (orange area) at $T=\SI{0.15}{\milli\second}$ and at the final time $T=\SI{1.09}{\milli\second}$ 
(upper and lower panels), initial and goal states probability distributions are shown as reference (dashed blue and red lines).\\
b) Optimal control fields for transformation times $T = \SI{1.09}{\milli\second}$ for $N = 1, 700$ and $7000$ atoms obtained via full CRAB optimization (respectively red, blue, and yellow line). 
Inset: Fourier spectrum of the optimal control field for $T=\SI{1.09}{\milli\second}$ and $N=700$ (blue solid line). The vertical lines correspond to single particle transitions from the ground state (red).}
\label{fig:QSL}
\end{figure*}


\section*{Fast manipulation of the motional state of a BEC on an atom chip}


The first system for which we demonstrate time-optimized driving is a one-dimensional (1D) Bose-Einstein condensate (BEC) of Rubidium 87 atoms on an atom chip, performed at the Technische Universit\"at Wien. Optimal control is used to perform coherent transfers between eigenstates of the transverse confining potential. 
Such transfers are tools for probing non-equilibrium quantum dynamics and studying decay processes from excited eigenstates, for example the emission of twin-atom beams~\cite{Buecker2011}. Furthermore, coherent manipulation of non-classical motional states allows performing interferometric sequences with these states~\cite{vanFrank2014}, and opens perspectives for quantum information operations. For the useful implementation of such operations, as well as to separate the timescales of state initialization and of decay processes, the duration of the optimal control transfers is key. We characterize here the QSL for a full population transfer from the ground to the first excited state, and implement experimentally the predicted shortest possible transfer which allows keeping high transfer efficiency. 

The atom chip used in this experiment consists of setups of micro-fabricated structures on a surface, generating magnetic fields to trap neutral atoms~\cite{reichel2011}. They can produce strongly confining potentials and allow for very precise manipulation of the atoms. These capacities \cite{reichel2011} are here exploited to produce a well-defined transversally anharmonic potential and to accurately displace the trapping potential along the anharmonic direction, following a trajectory designed by QOC. To realize transfers between the BEC motional states, the anharmonicity is necessary as it lifts the degeneracy between the levels and allows transfers to specific eigenstates of the trap or superpositions thereof. It also induces a coupling between center-of-mass motion and intrinsic motion of the BEC \cite{ott2003}. 
There is no trivial way to perform these transfers fast, due to the presence of interactions {and higher energy levels}. In order to constrain the dynamics into the two-level system formed by the initial and target states, the minimum duration {for a Rabi driving field --- a weak amplitude, sinusoidal displacement at the level spacing frequency ---} should be defined by the detuning between the level spacings, {which is on the order of $\SI{0.6}{\kilo\hertz}$. This simple picture is complicated by the interactions, which shift the energy levels and are also responsible for population transfers between eigenstates. The required driving time, obtained from numerical simulations, exceeds $\SI{9}{\milli\second}$. This is much longer than the timescale of interaction-induced decay into twin-atom beams, about $\SI{3}{\milli\second}$ for our typical atom number \cite{Buecker2011}. Therefore, although the initial and final states can be described in a two-level model, the time-optimal transfer trajectory is expected to involve higher motional states.} Designing the necessary complex driving fields thus requires the use of QOC. 

As long as the decay processes can be neglected, there is no coupling between the different axes of the potential and the steering dynamics can be described by considering only the transverse $y$-direction along which the potential is displaced. Thus, in a mean-field approach, the dynamics of the condensate can be described by an effective 1D Gross--Pitaevskii equation (GPE):
\begin{align}
\label{eq:gpe}
i\hbar\frac{\partial\psi(y,t)}{\partial t}&=\hat H_\text{gp}[\psi]\psi(y,t),\nonumber\\
\phantom{=}\hat H_\text{gp}[\psi]&=\left[-\frac{\hbar^2}{2m}\frac{\partial^2}{\partial y^2}+V(y,t)+g_\text{1D}N\vert\psi(y,t)\vert^2\right],
\end{align}
with $N$ being the {number of atoms in the quasi-condensate} and $g_\text{1D}=g_\text{1D}(N)$ an effective 1D coupling constant {for the displacement direction $y$} (see appendix B)~\cite{pitaevskii_2003}. The potential along $y$ can be well approximated by 
$
\label{eq:Vy}
V(y,t)=p_2
\left[
y-\lambda(t)
\right]^2
+p_4\left[
y-\lambda(t)
\right]^4
+p_6
\left[
y-\lambda(t)
\right]^6,
$ 
where  $\lambda(t)$ is the control field and $p_2,\,p_4,\,p_6$ are fitting parameters of the trapping potential (see appendix C).


The system is initially prepared in the ground state $\phi_0(y)$ of the trap $V(y,0)$ with $N$ interacting bosons, and the target state is chosen as the corresponding first excited state $\phi_1(y)$. The relevant figure of merit is the infidelity  
$F_1=1 - \left\vert
\int_{\mathbb{R}}\mathrm{d}y\,\phi_1^*(y)\psi(y,T)\right\vert^2$,
where $\psi(y,T)$ is the final state of the system.

We perform a CRAB optimization including the limited bandwidth of the electronics and the maximum possible trap displacement $\lambda_\text{max}= \SI{1}{\um}$. The results for different transformation times $T$ are reported in Fig.~\ref{fig:QSL}. The infidelity $F_1$ decays monotonically 
with {one inflexion point, which we interpret as the crossover between} two characteristic timescales. 
Indeed, within the fastest timescale (about $\SI{0.15}{\milli\second}$) the optimal solution is to perform an almost rigid translation of the initial wave packet, which maximizes the overlap  with one of the two lobes of the first excited state of the trap (see upper inset in Fig.~\ref{fig:QSL}a). This results from the fact that simply displacing the initial ground state already yields a figure of merit of about \SI{60}{\percent}.
Solving the full problem obviously requires to modify the wavefunction shape (lower inset in Fig.~\ref{fig:QSL}a) by means of more complex manipulations of the system parameters, over a longer time. 
This optimal dynamics has also a geometric interpretation: it is composed by two optimal transitions, the first between the initial state and the intermediate one depicted in Fig.~\ref{fig:QSL}a (upper inset), and the second between the latter and the goal state. 
Each transformation displays a monotonic decay of the final figure of merit $F_1$ as a function of the transformation time $T$, which can be fitted 
via a $\cos^2(T)$ decay (blue lines in Fig.~\ref{fig:QSL}a), that is, they are compatible with two concatenated optimal transformations at the QSL. 
\begin{figure*}[t]
\centering
\begin{lpic}[l(0mm),r(0mm),t(0mm),b(0mm)]{Fig3a(86mm)}
\lbl[t]{2,60;{\fontfamily{phv} \selectfont \large a)}}
\lbl[t]{90,60;{\fontfamily{phv} \selectfont \large b)}}
\end{lpic}
\def\svgwidth{86mm}
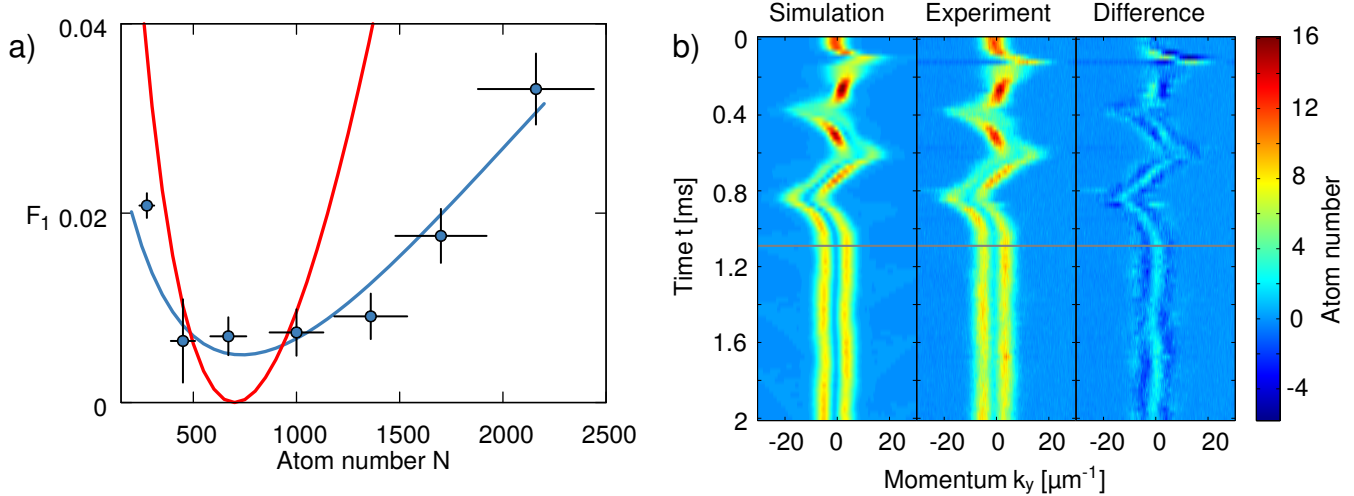
\caption{a) Theoretical predictions for the final infidelity $F_1$ as a function of the atom number when using the control fields optimized for 700 atoms. The shown numerical results are obtained for total transformation times $T=1.09, 5.01$ ms (blue and red lines) and are compared to experimental results (circles) obtained with the optimal control field for $T=1.09$ms.\\
b) Transverse distribution after time-of-flight during the optimal process ($t < T_\text{OPT}$) and after ($T_\text{OPT} < t < \SI{2}{\milli\second}$) for $N=700$: experiment (center), corresponding GPE simulation (left) and residual difference (right). The gray horizontal line highlights $T_\text{OPT}$.} 
\label{fig:results}
\end{figure*}

In summary, the optimal process taking into account experimental constraints and finite measurement precision (on the order of $\SI{1}{\percent}$, indicated by the green interval in Fig.~\ref{fig:QSL}a) lasts $T_\text{OPT}\simeq \SI{1.09}{\milli\second}$ and reaches an infidelity $F_1 \simeq \num{0.005}$. This time-optimal transfer is about five time shorter than previously achieved~\cite{Buecker2011, bucker2013}. The corresponding optimal control field for $N=700$ atoms is shown in Fig.~\ref{fig:QSL}b (blue line). This control field was used as an initial guess for optimizing two other pulses, for different atom numbers $N=1$ and $7000$. As shown in Fig.~\ref{fig:QSL}b, the resulting pulses have similar shape but with clear deviations. 

To investigate the composition of the optimal transfer control field and gain some physical insight into it, we performed a Fourier analysis of the optimal control field for $T_\text{OPT}$ and $N=700$ atoms, shown in the inset of Fig.~\ref{fig:QSL}b. Frequencies beyond $\sim$ 25 kHz do not play any relevant role. For lower frequencies, the spectrum has a rather continuous behavior with some prominent peaks. We compared the structure of the spectrum with the transitions of the single particle Hamiltonian (vertical lines in the 
inset). It appears that the main peaks are close to single-particle transitions from ground state to excited states, showing that many eigenstates of the potential are involved in the transfer dynamics. However, not all peaks could be clearly matched with a single physical transition, either single particle or a collective Bogoliubov excitation, see Eq. \ref{Eqn:Lbg} in appendix B (not shown in the inset of Fig. \ref{fig:QSL}). {This analysis emphasizes the complexity of the optimized control field and also the difficulty of engineering and understanding these optimized control fields in intuitive ways.}


The optimal control field obtained above, although promising, would be useless if it were not stable against experimental fluctuations.  
In the present experiment, atom number fluctuations are the main source of perturbation. In normal conditions, fluctuations of the order of $\SI{10}{\percent}-\SI{20}{\percent}$ of the atom number are unavoidable. Therefore, we checked the stability of the optimal process described above against fluctuations of the number of atoms $N$. 
The numerical results are reported in Fig.~\ref{fig:results}a, for different atom numbers and different transformation times $T$, one corresponding to the optimal time $T_\text{OPT}$ and another one about five times as long, comparable with the transformation time used in the experiment of Ref.~\cite{Buecker2011}. As can clearly be seen, the slower process results in a better theoretical infidelity $F_1$ at $N=700$; however its sensitivity against atom number fluctuations is much higher, due to the fact that the effects of atom interactions are integrated over a longer time. Eventually, for atom number fluctuations above $\SI{20}{\percent}$, the performances of the slower optimal protocols become even worse than those of the faster ones. 
{In order to further investigate the effects of atom interactions, we also simulated the application of the the optimal control field computed for $N=1$ to the interacting system. We obtained for the optimal time an infidelity $F_1 = 0.07 $, an effect that becomes much worse for the long pulse, with $F_1 = 0.25 $. This result reflects again the fact that the effects of interactions accumulate with time, but also that interactions must be taken into account to obtain good results, including at the optimal time.}

\begin{figure*}[t]
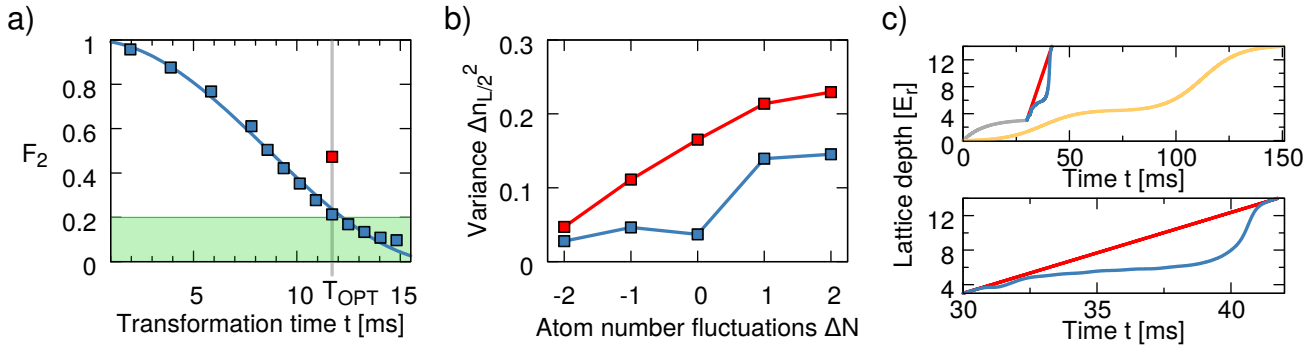

\centering
\begin{lpic}[l(0mm),r(0mm),t(0mm),b(0mm)]{Fig4a(57mm)}
\lbl[t]{2,45;{\fontfamily{phv} \selectfont \large a)}}
\end{lpic}
\begin{lpic}[l(0mm),r(0mm),t(0mm),b(0mm)]{Fig4b(57mm)}
\lbl[t]{2,45;{\fontfamily{phv} \selectfont \large b)}}
\end{lpic}
\begin{lpic}[l(0mm),r(0mm),t(0mm),b(0mm)]{Fig4c(57mm)}
\lbl[t]{2,45;{\fontfamily{phv} \selectfont \large c)}}
\end{lpic}
\caption{ a) Theoretical prediction of the optimal figure of merit $F_2$ as a function of the control field duration $T$ (blue squares). The blue solid line is a fit determining the QSL, the crossing point between the numerical result and the green region (estimated  experimental limitations) defines the optimal time $T_\text{OPT} = \SI{12 \pm 0.2}{\ms}$ (gray vertical line). The red square displays the final $F_2$ for the fast linear control field.\\
b) Theoretical prediction for the final atom number fluctuation in the center of the trap $\Delta n_{L/2}^2(T)$ for different atom numbers $N$ for the linear (red) and optimal (blue) control fields. 
The optimized control field works best for 16 atoms, but a small deviation only slightly decreases the figure of merit (blue). The linear ramp results in a higher figure of merit for all atom numbers (red).\\
c) Lattice ramps used in the experiment.  Top: the lattice is first slowly ramped to \SI{3}{\Er} (grey) before either the fast linear control field (red) or fast optimized control field (blue) is applied. The typical adiabatic control field (yellow) is much longer. 
Bottom: magnification of the comparison between the linear (red) and the optimal control field (blue).}
\label{fig:qsl}
\end{figure*}

\subsection*{Experiments}

We describe the experimental test of the optimal process engineered theoretically in the previous section. A BEC is prepared in the transverse ground state of an elongated potential. As illustrated in Fig.~\ref{fig:ls}b, the BEC is trapped and manipulated by current-carrying wires on an atom chip. At any point during the manipulation sequence or after it, the atomic ensemble can be released from the trap and imaged when it crosses a light sheet after a time-of-flight (for details on the experimental setup and sequence, see appendix C) \cite{bucker2009}. 
The profiles of the transverse atomic distribution after time-of-flight are then stacked to construct an image of the time evolution of the transverse momentum distribution during the transfer to the first excited state (Figure ~\ref{fig:results}b, middle panel).
Figure~\ref{fig:results}b shows a typical experimental result and comparison to simulations: the image in the middle represents the experimental transverse momentum distribution (fluorescence measurement results) as it evolves in time, starting from the beginning of the transfer field. This can be qualitatively compared with the GPE numerical simulation (left panel), or more quantitatively by plotting their difference (right panel).
The transfer efficiency is inferred from the evolution of the momentum distribution after the application of the control field, e.g. after $T_{\mathrm{OPT}} = \SI{1.09}{\milli\second}$. This distribution is fitted with Gross-Pitaevskii equation simulations, where the fit parameter of interest is the population in the first excited state.
Finally, this analysis yields a transfer efficiency of $\SI{99.3 \pm 0.6}{\percent}$, corresponding to an estimated figure of merit $F_1 =0.7 \% \pm 0.6 \%$, in excellent agreement with the theoretical prediction. We also repeated the experiment with the same control field for different atom numbers, obtaining a good agreement between theory and experiment over one order of magnitude of different atom numbers $N$, as shown in Fig.~\ref{fig:results}a.\\

We therefore demonstrated that applying QOC to the atom chip system provides a fast, robust, and efficient method for state initialization and manipulation. This fast state manipulation allowed performing a Ramsey interferometer with motional states \cite{vanFrank2014}. In the next section, we apply the same optimization algorithm to a different system, showing that the optimal steering demonstrated here is not dependent on this particular experimental setup or process, but it shall be expected in general.


\section*{Mott-insulating ground state of bosonic atoms in an optical lattice}


The phase transition between the superfluid (SF) and the Mott-insulating (MI) {phase in cold atoms} has been studied in different experiments~\cite{Georgescu2014}, and  nowadays the MI state with typically unit filling is used{, for instance,} as {a well-defined} initial state for {experiments simulating the dynamics of effective spin systems~\cite{hild2014, fukuhara2015, brown2015}}. These experiments start with a BEC and cross the phase transition to the desired MI state by slowly increasing the depth of the optical lattice. In an infinite-size homogeneous system these two states are separated by a quantum critical point and therefore cannot be adiabatically connected by varying the lattice depth.
In typical experimental systems, however, the finite number of atoms and the presence of an additional confining potential turn the phase transition into a crossover, thereby opening the possibility for an adiabatic preparation of the MI ground state. Here we demonstrate that a faster, non-adiabatic optimal steering across the 1D SF-MI crossover, is possible. The optimized control field we engineer speeds up most of the ramp of the system by a factor of ten compared to the adiabatic protocol, from \SIrange{120}{12}{\ms}, without measurable additional distortion of the final state.

The system we use is composed of parallel tubes containing on average 16 Rubidium 87 atoms each. An optical lattice of depth $V(t)$ is applied along these tubes. Each tube is described by the one-dimensional bosonic Hubbard Hamiltonian~\cite{Jaksch2005}
\begin{eqnarray}
	\hat H_{\text{BH}} &=& -J \sum_i \left( \hat b^\dag_i \hat b_{i+1} + \hat b^\dag_{i+1} \hat b_i \right) + \frac{U}{2} \sum_i \hat n_i (\hat n_i-1) \nonumber \\
	&&+ \frac{1}{2} m \omega^2 a_{\text{lat}}^2 \sum_i (i-i_0)^2 \hat n_i \; ,
	\label{ham}
\end{eqnarray}
where the index $i=1, \dots, L$ labels the lattice sites and $i_0=(L-1)/2$ the center of the trap, $J$ is the tunnel coupling between neighboring sites, $U$ is the on-site interaction strength,  $\omega$ denotes the harmonic confining potential, and $a_{\text{lat}}$ the lattice spacing. The operator $\hat b_i$ ($\hat b^\dag_{i}$) annihilates (creates) an atom at site $i$ while $\hat n_i = \hat b^\dag_i \hat b_i$ counts the number of atoms at that site. In the absence of the harmonic confinement ($\omega = 0$), the critical point of the SF-MI transition is located at $U/J \approx 3.4$~\cite{kuhner_2000}, corresponding to a lattice depth $V_\text{c} = \SI{4.5}{\Er}$, where $\si{\Er} = (2\pi\hbar)^{2}/(8 m a_{\text{lat}}^{2})$ is the recoil energy of the lattice and $m$ the atomic mass of the atoms. All the presented theoretical results are obtained in the presence of a trapping frequency $\omega=2\pi\times 63.5$ Hz equal to the frequency measured in the experimental setup, and unless stated otherwise, simulating $N = 16$ atoms in a lattice of $L=32$ sites. 

The dynamics of the Hamiltonian~\eqref{ham} is simulated via the time-dependent density matrix renormalization group algorithm (t-DMRG, see appendix D). The time-dependent tunnel coupling $J(t)$ and the on-site interaction energy $U(t)$ are derived from the lattice depth $V(t)$ by calculating the overlap integrals of the Wannier functions for the single-particle problem~\cite{Jaksch2005}. 
The bosonic Hubbard Hamiltonian, provided by Eq.~\eqref{ham}, is only a good description of the system for sufficiently large lattice depths (typically $V > \SI{3}{\Er}$). We therefore assume the system to be initialized at a lattice depth of \SI{3}{\Er} and we
{optimize the functional dependence in time of the lattice depth for a ramp ranging from \SI{3}{\Er} to \SI{14}{\Er}} driving the SF-MI crossover.

\begin{figure}[t]
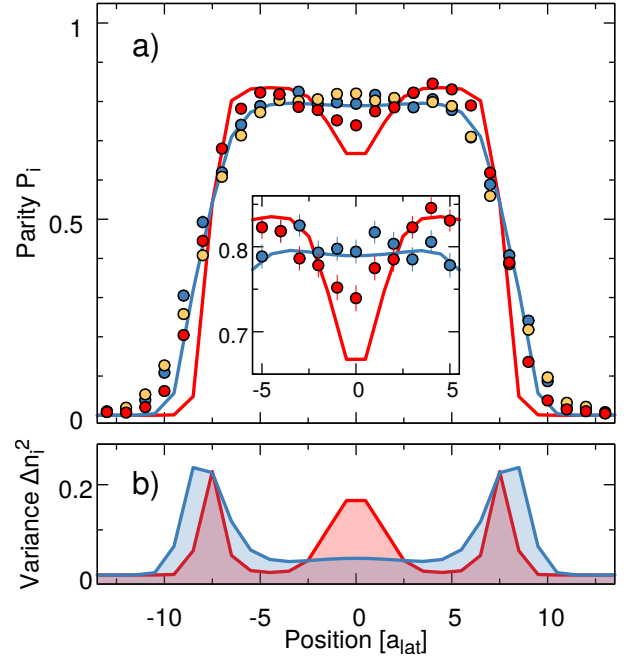

\begin{lpic}[l(0mm),r(0mm),t(0mm),b(0mm)]{Fig5(86mm)}
\lbl[t]{19,84;{\fontfamily{phv} \selectfont \large a)}}
\lbl[t]{19,26;{\fontfamily{phv} \selectfont \large b)}}
\end{lpic}
\caption{
a) Experimental mean parity profiles $P_i$ resulting from the adiabatic (yellow points), optimized (blue points) and linear (red points) lattice ramps compared to the rescaled numerical results (blue and red lines). The short linear lattice ramp has a dip in the parity profile at the center due to non-adiabatic effects. Inset: magnification of the central part of the main panel. Standard deviations of the measured data are smaller than data points in the main plots and therefore only shown in the inset.\\
b) The red and blue shaded areas display the numerically computed atom number fluctuations for the linear and the optimized ramps. 
}
\label{fig:par}
\end{figure}

The shape of the lattice depth $V(t)$ is optimized for different transformation times $T$ using the CRAB algorithm. The figure of merit we minimize is the 
rescaled average variance of the site occupancy in the center of the trap where we expect that the effect of the harmonic potential will be negligible
and we could observe a MI state. That is, $F_2(T) = \mfrac{1}{8} \sum_{i=L/2-4}^{L/2+4}  \Delta n_i^2(T)/\Delta n_i^2(0)$ where $\Delta n_i^2(t) = \langle \hat n_i^2(t) \rangle - \langle \hat n_i(t) \rangle^2$ and the sum runs over the eight sites at the center of the harmonic trap. The numerical figure of merit ranges from $F_2(T=0)=1$ to $F_2=0$ for a perfect Mott-insulator, while any residual excitations at the final time will increase $F_2(T)$. Figure~\ref{fig:qsl}a displays the resulting optimal figure of merit $F_2$ as a function of the transformation time $T$. As expected for an optimal crossing of a quantum phase transition~\cite{Caneva2009}, the numerical results are accurately described by the curve $F_2(T) = \cos^2(T/T_\text{QSL})$, resulting in $T_\text{QSL} = \SI{17.3 \pm 0.2}{\ms}$. 
However, due to experimental limitations arising from the non zero temperature of the 1D tubes (green region in Fig.~\ref{fig:qsl}a) the optimal verifiable figure of merit corresponds to an optimal time $T_\text{OPT} \sim \SI{12}{\ms}$.

Before proceeding with the experimental verification of the optimal process, we investigate the robustness of the optimal process with respect to total atom number fluctuations. 
We show in Fig.~\ref{fig:qsl}b the final density fluctuations at the center of the trap $\Delta n_{L/2}^2(T)$ under deviations $\Delta N$ of the atom number up to more than $10\%$ (i.e. $\Delta N \pm 2$) in the system for the optimized lattice ramp (blue). For comparison, we also show the result of a linear lattice ramp of same duration (red). 
As before, the optimal process displays a rather high level of robustness: the density fluctuations remain similar for all atom numbers, however larger atom numbers lead to a larger amount of defects in the density distribution. 
The corresponding optimal lattice ramp $V(t)$, together with a linear ramp and the adiabatic one for reference, is displayed in Fig.~\ref{fig:qsl}c.
The optimal protocol has been computed only for a system prepared in the ground state of the initial Hamiltonian, 
but we show below that the insensitivity to atom number fluctuations also provides a certain immunity against a finite initial temperature of the system.

\subsection*{Experiments}

The experimental implementation of the theoretically predicted optimal protocol presented in the previous section was performed on the quantum gas microscope experiment at the Max-Planck Institut f\"ur Quantenoptik in Garching. At the beginning a two-dimensional degenerate gas of polarized Rubidium 87 atoms is produced in a single anti-node of a vertical optical lattice (period $a_{\text{lat}} = \SI{532}{nm}$)~\cite{Sherson_2010, Endres2011} (see appendix E). {By slowly ramping up an additional optical lattice} along the $y$-axis, the system is divided into about 10 decoupled one-dimensional tubes (both lattices had a depth of \SI{20}{\Er}). A third optical lattice ($x$-axis), perpendicular to the other two, is used to drive the system from the SI to the MI phase by varying its depth $V(t)$ over time. The initial number of atoms is tuned to result in a lattice filling of one in the insulating phase, corresponding to about $N=16$ atoms in the central tubes.
At the end of this ramp, the density distribution is `frozen' by rapidly raising all three lattice depths to $\sim \SI{80}{\Er}$. Finally, the on-site parity projected atom density is detected by fluorescence imaging~\cite{Sherson_2010}.

The usual adiabatic crossing of the phase transition in two dimensions to the MI phase uses a double s-shaped ramp where the slope is minimum at the phase transition. Here we use a similar ramp for the one-dimensional systems as a reference point (see yellow line in the upper panel of Fig. \ref{fig:qsl}c). Each s-shaped ramp has a duration of \SI{75}{\ms} and the step is centered around the critical lattice depth of $V_\text{c}$. This adiabatic preparation leads to an average parity of the site occupancy of \num{0.80 \pm 0.01} in the center of the trap. This means that \SI{80}{\percent} of the sites are filled initially with one atom, the rest being either empty or filled with two atoms (the probability for a triple occupancy can be neglected). We attribute the \SI{20}{\percent} defects mostly to the finite temperature of the initial state (we note that in a 2D geometry average final parities of $>$\SI{96}{\percent} are typical) because this fraction does not significantly vary as we increase the duration of the lattice ramp.
The theoretically predicted optimal steering field is implemented in the experimental setup by increasing the lattice depth first from 0 to $V_\text{i}=\SI{3}{\Er}$ over \SI{30}{\ms} 
and then from $V_\text{i}$ to $V_\text{f}=\SI{14}{\Er}$ over \SI{11.75}{\ms} following the optimal control field. In order to reduce the atom number fluctuations in the experimental sample, we  restrict the analysis to the central 5 tubes in two-dimensional samples having a diameter of 16 lattice sites (see appendix E). 

The measured parity of the site occupancy in the final state is plotted in Fig.~\ref{fig:par}, where we compare the result of the optimal protocol to those of the adiabatic state preparation and of a linear fast control field of the lattice performed in \SI{11.75}{\ms}. We observe no significant difference between the optimal and adiabatic protocols (yellow and blue data sets): the optimal control technique therefore leads to a factor of ten speed up of the state preparation without loss of fidelity. The performance of the optimized protocol is better appreciated when comparing with the final state reached using a linear control field of the same duration. Here the parity of the site occupancy displays a dip in the center, surrounded by narrow bands of high parity. Such a distribution indicates that the redistribution of atoms across the system that is necessary to build an extended insulating region could not take place. Instead, insulating regions that formed locally around the points where the initial site occupancy was close to one have confined the excess particles in the center of the trapping potential~\cite{Bernier2011}.
Note that, as displayed in the inset of Fig.~\ref{fig:par},  the difference between the optimized and the linear control field  --- especially in the center of the trap --- differs by more than three standard deviations.  
Moreover, in Fig.~\ref{fig:par} we compare the experimental results with the numerical simulations for the optimal control fields: both experimental and numerical results feature a flat top profile of comparable width, but the maximum parity of the site occupancy achieved in the experiment yields $P_{\max} =\num{0.80\pm 0.01}$, compared to $P_{\max} =\num{0.96\pm 0.01}$ in the numerical simulation. This difference is a consequence of the simulation being performed at zero temperature, whereas the experimental system has a finite initial temperature. However, if rescaled by the experimental value of $P_{\max}$, there is a good agreement between them, that is, their difference can be parametrized with a single parameter corroborating the fact that their discrepancy originates most likely from the different temperature (purity) of the states.


\section*{Conclusions}

We theoretically engineered and experimentally implemented two fast optimally controlled processes for different paradigmatic complex cold atom systems:
the optimal preparation of a non-classical motional state of a BEC in a magnetic trap and the 1D superfluid-to-Mott-insulator crossover of cold atoms in an optical lattice. In particular, the latter is the first experimental demonstration of optimal control of a crossover related to a quantum phase transition in a finite system.
We demonstrated that optimal processes can be engineered and implemented for many-level systems with non-linearities as well as for many-body quantum systems, and that the theoretical estimate of the QSL can be experimentally reached.
We have shown numerically that the optimal processes are robust with respect to experimental imperfections, stable against atom number fluctuations (that are unavoidable without post-processing of the data) and finite temperature corrections, paving the way to a systematic utilization of optimal control in experiments with quantum many-body systems.

The optimal preparation of excited states of cold atoms on an atom chip, performed with an unprecedented fast process, open new perspectives for the development of accurate and sophisticated protocols for sensing, interferometry and cold atom manipulations. 
The numerical and experimental results on the SF-MI crossing demonstrate that the purity of the state reached by the fast optimal protocol is the same as the one obtained by means of the adiabatic protocol. This experiment indicates that, along the same lines, the generic adiabatic quantum computation scheme can be in principle performed in a fast and optimal way (i.e. not adiabatically). 
Finally, the speedup of these processes naturally reduces the detrimental effects of decoherence in the system and thus paves the way to the experimental realization of protocols of increasing complexity in the near future. 
~\\

The Garching team acknowledges support by T. Fukuhara, M. Endres, P. Schauß and J. Zeiher. The Vienna team acknowledges the support of R. Bücker, T. Berrada, J.-F. Schaff, M. Pigneur, M. Maiwöger, Wu RuGway and T. Schumm at different stages of the project.
A. N. acknowledges A. I. Streltsov for using the code of the multi-configurational time-dependent Hartree Method for bosons in order to check the validity of the mean-field description of the atom chip optimal process.
We acknowledge support from the German Research Foundation (DFG) through the SFB/TRR21, the Hamburg Centre for Ultrafast Imaging - Structure, Dynamics and Control of Matter at the Atomic Scale, the Max-Planck-Society, and the EU through SIQS, RYSQ, UQUAM, ETAB. This work was performed on the computational resource bwUniCluster funded by the Ministry of Science, Research and Arts and the Universities of the State of Baden-Württemberg, Germany, within the framework program bwHPC.


\section*{Appendix}

\subsection{CRAB optimal control}

Optimal control theory is devoted to find the solution to functional minimizations of the form $\min_{V(t)} F(V(t))$, where $V(t)$ is the control field and $F$ a figure of merit to be computed via a dynamical law that describes the time evolution of the system. In QOC problems, the dynamical law is given by a Liouvillian equation for the system density matrix, which for pure states reduces to the time-dependent Schr\"odinger equation. Typical figures of merit are the overlap fidelity of the final state with respect to some given target state, the final energy of the system or some other interesting properties of the final state or of the path followed between the initial and the final state. Finally, figures of merit might include also constraints as the maximal power used to drive the system, the limited band-width of the control field or any other experimental constraints to be satisfied. 
In this work we employ the CRAB optimal control approach, that is, the optimization is implemented by looking for an optimal control field of the form $V(t) = V_0(t)  \,f(t)$, where $V_0(t) $ is some guess function we can use to include our physical intuition on the problem, or a preexisting solution to a simplified version of the complete optimal control problem, and $f(t)$ is a correction expressed in a truncated and randomized basis. For example, one could work in a truncated Fourier series of the form 
$f(t) = 1+ \sum_{k}  [A_k \sin(\nu_kt) + B_k \cos(\nu_kt)]/\Gamma(t)$ where $k = 1, . . . ,n_f$, $\nu_k = 2\pi (k+ r_k)/T$ are randomized Fourier harmonics with $T$ the total time evolution, $r_k \in [-1/2,1/2]$ are random uniformly distributed, and $\Gamma(t)$ is a normalization function to keep the initial and final control field values fixed. 
The optimization problem is then reformulated as the extremization of a multivariable function $F({A_k}, {B_k}, {\nu_k)}$, which can be performed with standard numerical algorithms, also gradient-free to improve their efficiency~\cite{caneva_speeding_2011}.
This approach has been successfully applied to different systems and protocols and it has been shown that it allows to achieve an efficient control of many-body quantum system dynamics~\cite{caneva_2014,doria_2011,rosi_2013,caneva_2012}. 
It has also been theoretically shown that the minimal value of the final figure of merit drops exponentially with the number of optimization parameters $n_f$, property that guaranties in most cases of interests an efficient and quick convergence to optimal control fields that results in errors comparable to experimental ones~\cite{lloyd_2014}.

\subsection{Effective one-dimensional Gross-Pitaevskii equation}

The displacement of the trap, needed to excite a quasi-1D BEC as discussed in the first experiment of the paper, occurs along one of the two transverse directions, where the confinement is much stronger than in the axial direction (the frequency ratio between the transverse and the axial confinements is about $10^2$). Usually, see for instance Ref.~\cite{pitaevskii_2003}, a quasi-1D GPE for the axial motion is derived under the assumptions that the motion is effectively frozen to the ground state of the transverse confinement and that the atomic interactions can be described by a contact potential. Nonetheless, we shall now show that an effective GPE for the motion along one of the two transverse directions (hereafter the $y$ axis) can be obtained, since the dynamics along the $y$-direction is much faster than in the axial one (hereafter the $x$ axis), and therefore phononic (axial) excitations can be neglected during the transfer process.

To this end, we first compute the Heisenberg equation of motion for the atomic quantum field operator $\hat\Psi(\mathbf{r})$. Then, we perform the replacement 
$\hat\Psi(\mathbf{r} )\mapsto\Psi(\mathbf{r} )=\sqrt{N}\psi(x)\psi(y)\psi(z)$ with $N$ being the atom number, $\psi(x)=\sqrt{n_1(x)}$, $\psi(z)=\left(\frac{m\omega_z}{\pi \hbar} \right)^{1/4}e^{-\frac{m\omega_z}{2\hbar}z^2}$, whereas the axial atomic density of the quasi condensate is given by
$n_1(x)=\alpha [1-(x/L)^2]\{\alpha [1-(x/L)^2]+4\}/(16a_{\mathrm{3D}}^{\mathrm{s}})$
such that $\int_{-L}^L\mathrm{d}x\,n_1(x) = N$~\cite{Gerbier2004} ($a_{\mathrm{3D}}^{\mathrm{s}}$ is the three dimensional s-wave scattering length). 
Here $m$ is the atomic mass, $\omega_z$ the trap frequency of the harmonic confinement in the $z$ direction, 
2$L = 2a_x^2\sqrt{\alpha}/a_{\perp}$ the size of the condensate along the axial direction with 
$a_x=\sqrt{\hbar/m\omega_x}$, $a_{\perp}=\sqrt{\hbar/m\omega_{\perp}}$, and 
$\omega_{\perp}=\sqrt{\omega_y\omega_z}$. The parameter $\alpha$ is obtained by solving the algebraic equation
$\alpha^3(\alpha+5)^2-(15 Na_{\perp}a_{\mathrm{3D}}^{\mathrm{s}})^2/a_x^4=0$.
Now, by integrating over $x$ and $z$ the equation of motion of the matter-wave field $\Psi(\mathbf{r})$, we obtain the effective GPE reported in Eq.~\eqref{eq:gpe}, 
with $g_{\mathrm{1D}}(N)=g_{\mathrm{3D}}\mathcal{I}_x/(\sqrt{2\pi}a_z)$, $g_{3D} = 4 \pi\hbar^2 a_{3D}^{\mathrm{s}}/m$, and 
$\mathcal{I}_x=\alpha^2 L (\alpha^2+9\alpha+21)/[315 (a_{3D}^{\mathrm{s}}N)^2]$.
Hence, contrary to the usual coupling constant for a quasi-1D trapped atomic Bose gas~\cite{Olshanii1998}, in our scenario the coupling constant relies on the atom number as well. We also note that for a fixed atom number the nonlinearity in Eq.~(\ref{eq:gpe}) is smaller than in a genuine quasi-1D quantum Bose gas, and therefore a multiorbital description of the dynamics does not provide any significant improvement to our mean field theory, as we have checked via the multi-configurational time-dependent Hartree method for bosons~\cite{Alon2008}.

Finally, we note that in order to analyze the structure of the spectrum of the optimal control field, we have solved the Bogoliubov--de Gennes equations~\cite{pitaevskii_2003} for the GP ground ($\phi_0$) and first excited ($\phi_1$) states, that is, we solved the eigenvalue equations $\mathcal{L}_{0,1}(u_k,v_k)^T = \hbar\omega_k (u_k,v_k)^T$, with the Bogoliubov--de Gennes operator given by~\cite{Castin1998}

\begin{align}
\label{Eqn:Lbg}
\mathcal{L}_{0,1} &= \left(
\begin{array}{cc}
\hat H_{\rm gp}[2 |\phi_{0,1}|^2 ] - \mu_{0,1} & \hspace*{-0.5em}
g_{\mathrm{1D}}  \phi_{0,1}^2 \\
-g_{\mathrm{1D}}  \phi_{0,1}^{*2} & 
\hspace*{-0.5em}
- \hat H_{\rm gp}[2 |\phi_{0,1}|^2] + \mu_{0,1}
\end{array}
\right).
\end{align}
Here $\mu_0\,(\mu_1)$ is the chemical potential corresponding to $\phi_{0}( y )$ [to $\phi_{1}( y )$]. For more details, we refer to Ref.~\cite{Castin1998}.

\subsection{BEC on atom chip experimental setup}

The experimental setup consists in a quantum degenerate Bose gas of Rubidium 87 atoms trapped on an atom chip. The atom chip is a square multilayer structure covered in current-carrying gold wires. The central DC wire together with homogeneous external magnetic fields form a strongly confining anisotropic Ioffe-Pritchard trap of aspect ratio of $\sim 200$. Transversally, the trap is dressed by radio-frequency fields to form an effective slightly anharmonic potential~\cite{Lesanovsky2006}.
As we outlined in the main text, the exact anharmonic potential has been approximated with a polynomial $V(y,t) = p_2
\left[
y-\lambda(t)
\right]^2
+p_4\left[
y-\lambda(t)
\right]^4
+p_6
\left[
y-\lambda(t)
\right]^6$, whose fit parameters are: 
$p_2 = 2\pi \hbar \times 310/r_0^2 \, \SI{}{J/m^2}$, 
$p_4 = 2\pi \hbar \times 13.6/r_0^4 \, \SI{}{J/m^4}$, and 
$p_6 = 2\pi \hbar \times 0.0634/r_0^6 \, \SI{}{J/m^6}$ with $r_0=172\cdot 10^{-9}$m.
This experiment was performed with the potential described in Ref.~\cite{vanFrank2014}.
The measured frequency in the $y$ direction is $\nu= \SI{1.77}{\kilo\hertz} $. The atomic gas is cold enough ($T < \SI{50}{\nano\kelvin} \simeq h/k_{\text{B}} \times \SI{1}{\kilo\hertz}$) and the chemical potential small enough ($\mu/h \simeq \SI{0.6}{\kilo\hertz}$ for $N \sim 700$) that the system sits in the ground state of this potential.

To realize transfers between motional states, the potential is displaced according to the optimized control field using an external wire, located as far away from the trapping wires as possible. This simple scheme leads to a close-to-horizontal displacement, with, however, a $\SI{19}{\degree}$ tilt with respect to the $y$ direction. Excitations in the $z$-direction are anyway limited by the anisotropy of the potential but, to avoid them more completely, the angle can also be compensated for by tilting the axes of the trapping potential using the radio-frequency dressing. This configuration has been used to take the present data. However, an alternative scheme using an offset current on the radio-frequency wires has been implemented as well, enabling a purely horizontal displacement and leading to similar results.

Following the displacement in real time is not possible, but recording the evolution of the momentum distribution in the trap is. For this, the atomic cloud is released from the trap at different times during and after the transfer process, and imaged after $\SI{46}{ms}$ time of flight. These images are integrated to obtain the density distribution along the $y$ direction as illustrated on Fig. 1b, and stacked to construct and represent the time evolution of the density after time-of-flight as shown on Fig. 3b. 
Due to the fast expansion of the cloud after the strongly confining trap is switched off, the interactions become rapidly negligible and the imaged density is homothetic to the in-trap momentum distribution. The atomic cloud is imaged by fluorescence when it falls through the light sheet, which is a very thin ($\sim \SI{40}{\micro\meter}$) layer of laser light formed by two counter-propagating laser beams, slightly detuned from resonance. Part of the emitted photons are then captured by an objective located below the light sheet, and the atom imaged with an EMCCD camera~\cite{bucker2009}.

The efficiency of the transfer and the errorbars on it are extracted from these density evolution patterns by comparison to GPE simulations, as in Ref.~\cite{vanFrank2014}. 
The 1D GPE evolution of the momentum distribution in the transverse potential is calculated starting from an initial superposition of states $ | \Psi_{\text{initial}} \rangle = \sum_k \sqrt{p_k} e^{i \theta_k} | k \rangle $ where $ k \in {0,1,2}$ corresponds to the ground, first excited and second excited states in the transverse direction of interest. A fit to the experimental data with the GPE simulation of the momentum distribution evolution, using the $p_k$'s and $\theta_k$'s as fit parameters, yields the ratio of population in the first excited state, $p_1$, which corresponds to the fidelity as defined for the optimization.

\subsection{Density matrix renormalization group}

The Density Matrix Renormalization Group (DMRG) is a numerical technique tailored to one-dimensional, correlated quantum many-body systems on a lattice~\cite{white_1992,schollwock_2011}.
In its modern formulation, it exploits a tensor-network ansatz (namely a Matrix Product State) -- dependent on an auxiliary dimension $\chi$ --
to efficiently represent a many-body state with a polynomial number of free parameters as a function of the number of lattice sites.  By means of different minimization techniques it is possible to obtain a faithful representation of the eigenstate properties of the system and exploiting the few-body short-range nature of the interaction among different sites and the Suzuki-Trotter formalism, it is possible to numerically simulate the time evolution of many-different setups. In particular, the process considered here can be efficiently simulated with up to tens of sites $L$ and particles 
$N$.
Recently, t-DMRG has been merged with optimal control theory by means of the CRAB optimal control technique described in appendix $A$. 
All simulations have been made with auxiliary dimension of up to $\chi= 24$, Trotter step $\Delta t= 10^{-2}\hbar/E_{\mathrm{r}}$ and truncation error below $\epsilon \le 10^{-5}$.


\subsection{Cold atoms in the optical lattice experimental setup}

For preparing a Rubidium 87 gas in a 1D geometry, we first prepare a two-dimensional (2D) gas by loading a three-dimensional BEC in a red-detuned optical lattice along the vertical $z$ direction, which is the imaging direction. We then single out one of the filled lattice nodes and remove all the others by combining a strong magnetic field gradient along the lattice axis, a microwave transfer between two hyperfine states and a resonant optical pulse, as described in Ref.~\cite{Sherson_2010}. Once a single 2D gas is isolated, we adiabatically switch on (\SI{200}{\ms}) a second optical lattice in the horizontal $y$ direction, thereby forming an array of 1D tubes. The remaining tunnel coupling strength between neighboring tubes is estimated to be $J/h=\SI{5.05}{Hz}$ using standard band structure calculations, which has a negligible effect over the duration of the experiment. The lattice along the 1D gas ($x$-axis), which drives the transition from the superfluid to the Mott-insulating state, is turned on and slowly ramped up to \SI{3}{\Er} within \SI{30}{\ms}. This is the initial state for the ramps to \SI{14}{\Er} described in the main text.

Due to the presence of an external harmonic confinement in each direction of space, the length of the 1D gas is bound to at most 16 sites in order to maintain a filling of one atom per site in the Mott-insulating regime. For the data analysis we focused on the central 5 tubes so as to ensure that the harmonic confinement along the tubes was approximately constant. The experiment was repeated 176 times for each control field profile (linear, adiabatic and optimal), therefore yielding a statistical ensemble of 880 independent samples. Because the imaging process only gives access to the parity of the density distribution we could not sort the samples according to their atom number. We therefore included all available data in the analysis. However, based on the measured radius of the 2D gas, we are confident that the atom number in the central 5 tubes was close to 16 in average with fluctuations of the order of $\pm 1$ atom. The temperature of the gas in the Mott-insulating regime can be extracted from the measured density profile~\cite{Sherson_2010}. We find typically $T \sim 0.1~U/k_\text{B}$, where $k_\text{B}$ is the Boltzmann constant. We were unable to achieve lower temperatures with the protocol used in this study, where the inital 2D gas is first loaded in the optical lattice along the $y$-direction (instead of the $x$- and $y$-lattices to be ramped simultaneously).








\bibliography{bibliography}



\end{document}

%% file: Results_colomap_v1.pdf_tex
\begingroup%
  \makeatletter%
  \providecommand\color[2][]{%
    \errmessage{(Inkscape) Color is used for the text in Inkscape, but the package 'color.sty' is not loaded}%
    \renewcommand\color[2][]{}%
  }%
  \providecommand\transparent[1]{%
    \errmessage{(Inkscape) Transparency is used (non-zero) for the text in Inkscape, but the package 'transparent.sty' is not loaded}%
    \renewcommand\transparent[1]{}%
  }%
  \providecommand\rotatebox[2]{#2}%
  \ifx\svgwidth\undefined%
    \setlength{\unitlength}{463.23759766bp}
    \ifx\svgscale\undefined%
      \relax%
    \else%
      \setlength{\unitlength}{\unitlength * \real{\svgscale}}%
    \fi%
  \else%
    \setlength{\unitlength}{\svgwidth}%
  \fi%
  \global\let\svgwidth\undefined%
  \global\let\svgscale\undefined%
  \makeatother%
  \begin{picture}(1,0.73372)
{

\fontfamily{phv}
\fontsize{0.33cm}{1em}
\selectfont

    \put(0,0){\includegraphics[width=\unitlength]{Results_colomap_v1.pdf}}%
    \put(0.33577889,0.03226271){\makebox(0,0)[lt]{\begin{minipage}{0.23377189\unitlength}\raggedright Momentum~k$_{\text{y}}$~[\textmu m\textsuperscript{-1}]\end{minipage}}}%
    \put(0.01166045,0.29371247){\rotatebox{90}{\makebox(0,0)[lt]{\begin{minipage}{0.0394185\unitlength}\raggedright Time~t~[ms]\end{minipage}}}}%
    \put(0.15750388,0.05698514){\makebox(0,0)[lb]{\smash{-20}}}%
    \put(0.25200115,0.05698514){\makebox(0,0)[lb]{\smash{0}}}%
    \put(0.32764145,0.05698514){\makebox(0,0)[lb]{\smash{20}}}%
    \put(0.10422821,0.67769885){\makebox(0,0)[lb]{\smash{0}}}%
    \put(0.0709785,0.56206389){\makebox(0,0)[lb]{\smash{0.4}}}%
    \put(0.0709785,0.20990343){\makebox(0,0)[lb]{\smash{1.6}}}%
    \put(0.10422821,0.08716979){\makebox(0,0)[lb]{\smash{2}}}%
    \put(0.07272742,0.326095){\makebox(0,0)[lb]{\smash{1.2}}}%
    \put(0.07174621,0.44044873){\makebox(0,0)[lb]{\smash{0.8}}}%
    \put(0.96361368,0.24705857){\makebox(0,0)[lb]{\smash{0}}}%
    \put(0.96672534,0.35752276){\makebox(0,0)[lb]{\smash{4}}}%
    \put(1.04,0.251268113){\rotatebox{90}{\makebox(0,0)[lb]{\smash{Atom number}}}}%
    \put(0.96718839,0.4678395){\makebox(0,0)[lb]{\smash{8}}}%
    \put(0.96512321,0.57222784){\makebox(0,0)[lb]{\smash{12}}}%
    \put(0.96627457,0.68405381){\makebox(0,0)[lb]{\smash{16}}}%
    \put(0.95838734,0.13970603){\makebox(0,0)[lb]{\smash{-4}}}%
    \put(0.40618297,0.05678149){\makebox(0,0)[lb]{\smash{-20}}}%
    \put(0.65356054,0.05678149){\makebox(0,0)[lb]{\smash{-20}}}%
    \put(0.39979449,0.74791316){\color[rgb]{0,0,0}\makebox(0,0)[lt]{\begin{minipage}{0.14158087\unitlength}\raggedright Experiment\end{minipage}}}%
    \put(0.15862913,0.74791316){\color[rgb]{0,0,0}\makebox(0,0)[lt]{\begin{minipage}{0.12757837\unitlength}\raggedright Simulation\end{minipage}}}%
    \put(0.65892511,0.74791316){\color[rgb]{0,0,0}\makebox(0,0)[lt]{\begin{minipage}{0.12602253\unitlength}\raggedright Difference\end{minipage}}}%
    \put(0.50249606,0.05698514){\makebox(0,0)[lb]{\smash{0}}}%
    \put(0.57486982,0.05698514){\makebox(0,0)[lb]{\smash{20}}}%
    \put(0.75180898,0.05660111){\makebox(0,0)[lb]{\smash{0}}}%
    \put(0.82763669,0.05660111){\makebox(0,0)[lb]{\smash{20}}}%

}
  \end{picture}%
\endgroup%